\pdfoutput=1
\documentclass[aps,prl,twocolumn,superscriptaddress]{revtex4}
\usepackage{graphicx}
\usepackage{amsmath}
\usepackage{color}

\begin{document}

\title{Coherent and incoherent metamaterials and the order-disorder transitions}

\author{N. Papasimakis}
\affiliation{Optoelectronics Research Centre, University of
Southampton, SO17 1BJ, UK}

\author{V. A. Fedotov}
\affiliation{Optoelectronics Research Centre, University of
Southampton, SO17 1BJ, UK}

\author{Yuan Hsing Fu}
\affiliation{Optoelectronics Research Centre, University of
Southampton, SO17 1BJ, UK}
\affiliation{Department of Physics,
National Taiwan University, Taipei 10617, Taiwan}

\author{Din Ping Tsai}
\affiliation{Department of Physics, National Taiwan University,
Taipei 10617, Taiwan}

\author{N. I. Zheludev}
\email{niz@orc.soton.ac.uk} \homepage{www.nanophotonics.org.uk/niz}
 \affiliation{Optoelectronics
Research Centre, University of Southampton, SO17 1BJ, UK}


\date{\today}

\begin{abstract}
We demonstrate a new class of "coherent" metamaterials, where a
regular ensemble of meta-molecules shows a collective, i.e.
coherent, narrow band resonant response, while disordering the
ensemble leads to broadening and eventually disappearance of the
resonance. We draw parallels between the observed collective
behavior of meta-molecules and the M\"{o}ssbauer effect and notice
certain remarkable similarities with the phase transitions of
ferromagnetic systems.
\end{abstract}

\maketitle

Artificial electromagnetic metamaterials provide a uniquely fertile
ground for achieving all kinds of unusual functionalities: they show
a negative index of refraction required for the creation of
diffraction-free superlenses \cite{pendry}, exhibit strong optical
magnetism \cite{soukoulis, shalaev} and impose asymmetric
transmission of light \cite{nanoassy}. Metamaterial structures can
be invisible \cite{fs}, act as electromagnetic cloaks \cite{cloak1,
cloak2}, show exceptionally high or zero refractive indices
\cite{mwall} and even behave like optical frequency
"superconductors" repelling the magnetic field of the optical wave,
thus mimicking the Meissner effect \cite{mwallopt}. The recent
development of self-assembly techniques for fabrication of
metamaterials that yield randomized arrays of meta-molecules
\cite{dorota} and new ideas for using metamaterials in coherent
sources of electromagnetic radiation \cite{spaser} stimulated our
interest in the effects of positional disorder on the
electromagnetic properties of two-dimensional metamaterial arrays.
In this work we have identified two distinctively different classes
of artificial structures, the "coherent" and "incoherent"
metamaterials with narrow-band resonant spectral response. The
response of incoherent metamaterials is essentially determined by
the properties of the individual meta-molecules and is virtually
insensitive to positional disorder. In "coherent" metamaterials,
external electromagnetic excitation induces a magnetic response in
individual meta-molecules. Here the meta-molecular disorder
dramatically modifies the spectral response and the magnetization of
the structure exhibits a collective, i.e. "coherent" nature
underpinned by interactions between the magnetic moments of the
meta-molecules.
\begin{figure}[hb]
\includegraphics[width=80mm]{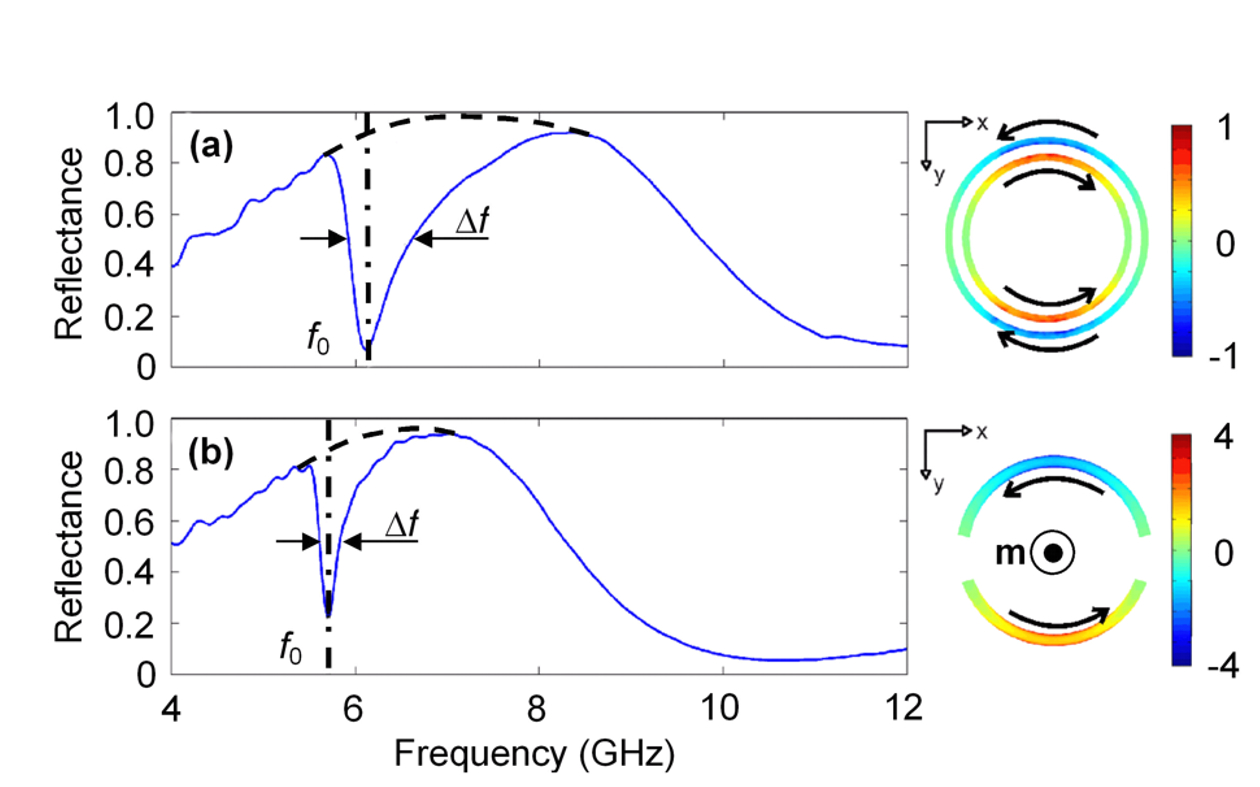}
\caption{ Normal incidence reflectivity spectra of regular
metamaterial arrays composed of pairs of concentric rings (a) and
asymmetrically-split rings (b), measured. The insets show the
induced anti-symmetric current modes corresponding to the
reflectance minima. In the case of the asymmetrically-split rings,
the anti-symmetric current configuration leads to a magnetic dipole
moment ($m$) normal to the plane of the rings. On the contrary,
concentric rings do not exhibit a magnetic moment at resonance.}
\end{figure}

Metallic ring meta-molecules are probably the most popular building
blocks of metamaterial structures. They are employed to provide a
strong magnetic response with negative permeability, required for
achieving a negative index of refraction. For this purpose, both the
ring and the wave propagation direction lie on the same plane, so
that the circular current induced on the ring creates a magnetic
moment \emph{parallel} to the magnetic field of the incident wave.
Here we investigate two novel types of metallic planar ring
metamaterials supporting electromagnetic modes with high quality
factors (see Fig.~1), where the magnetic moments associated with the
induced currents are \textit{perpendicular} to the plane of the
array. Meta-molecules of the first type were formed by pairs of
concentric metal rings, while meta-molecules of the second type had
the form of asymmetrically-split rings. In both types of
metamaterials studied in this work, the incident wave is normal to
the plane of the structure and therefore the interaction of the
induced magnetic moments with the magnetic field of the incident
wave is negligible. This ensures a key role of inter-meta-molecular
interactions in the formation of the electromagnetic response of the
material.
\begin{figure}[hb]
\includegraphics[width=80mm]{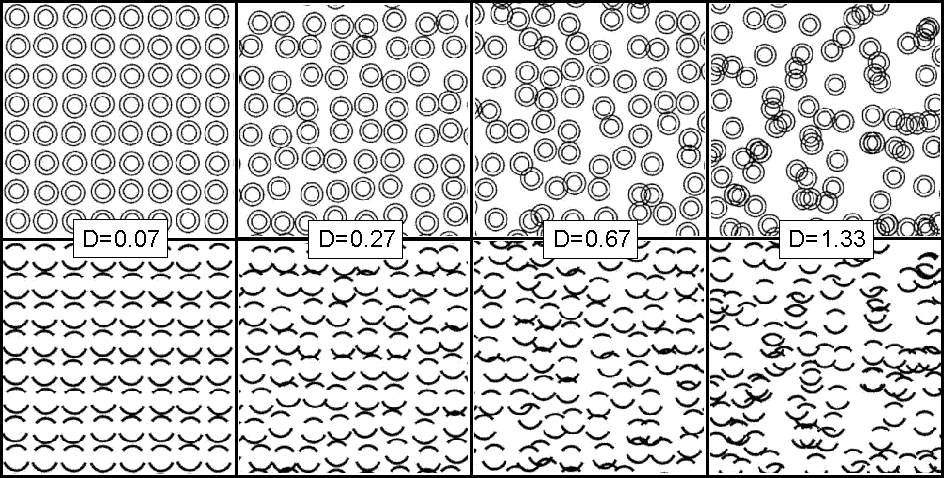}
\caption{Fragments of disordered metamaterial samples. The disorder
parameter, $D$, increases from left to right.}
\end{figure}

 A special feature of the studied structures is that they
support high-quality resonances associated with anti-symmetric
currents. In the asymmetrically split ring, these currents flow in
the longer and shorter arc of the ring, while in the concentric
rings they oscillate in the neighboring sections of the two rings
(see insets to Figs.~1a \& 1b). In both cases, the far-field
electric dipole radiation is cancelled due to the induced
counter-propagating currents, ensuring low radiation losses and thus
high-Q factors of the resonances \cite{asr}. Figure 1 shows
experimentally measured reflectance spectra of \emph{regular}
two-dimensional arrays for both types of meta-molecules. The radius
of the asymmetrically-split ring structure was $6~mm$ and the line
thickness $0.8~mm$, while the length of the arcs corresponded to
angles of 160 and 140 degrees. The radii of inner and outer rings
forming pairs of the concentric rings were $4.5~mm$ and $5.45~mm$
respectively and the line thickness of each ring was $0.4~mm$. The
unit cells of the regular structures had the size of $15 \times
15~mm$, which ensured no diffraction at frequencies below $20~GHz$.
In each case, the reflectivity spectrum features a broad background
of high reflection that is split by a narrow resonant dip associated
with high-quality anti-symmetric current modes.

 Despite the obvious similarity in the origin of the narrow
resonant response of the two metamaterials, their behavior becomes
dramatically different upon disordering the regular meta-molecule
lattice. The effect of disorder was studied experimentally by
introducing random displacements of the ring meta-molecules from the
perfect double-periodic grid according to a random uniform
distribution defined in the square interval ${x\epsilon
(-\alpha/2,\alpha/2),y\epsilon (-\alpha/2,\alpha/2)}$, while all
other parameters of the structures were fixed (see representative
cases in Fig.~2). A disorder parameter, $D$, was defined as the
ratio of $\alpha$ over the unit cell side. The evolution of the
reflectance spectra with increasing disorder is presented in Fig.~3.
For the concentric ring metamaterial, the resonant dip in the
reflectivity robustly retains its magnitude and width even for very
high levels of disorder, which may be seen as the persistence of the
blue band centered at $6.2~GHz$. For the asymmetrically-split ring
metamaterial, however, the similar resonant dip at $5.7~GHz$
degrades rapidly and completely vanishes for a moderate degree of
disorder. This is further illustrated in Fig.~3 with two
characteristic cross-sections of the spectral evolution
corresponding to cases of weak ($D_1 = 0.07$) and moderate ($D_2 =
0.40$) disorder.
\begin{figure}[ht]
\includegraphics[width=80mm]{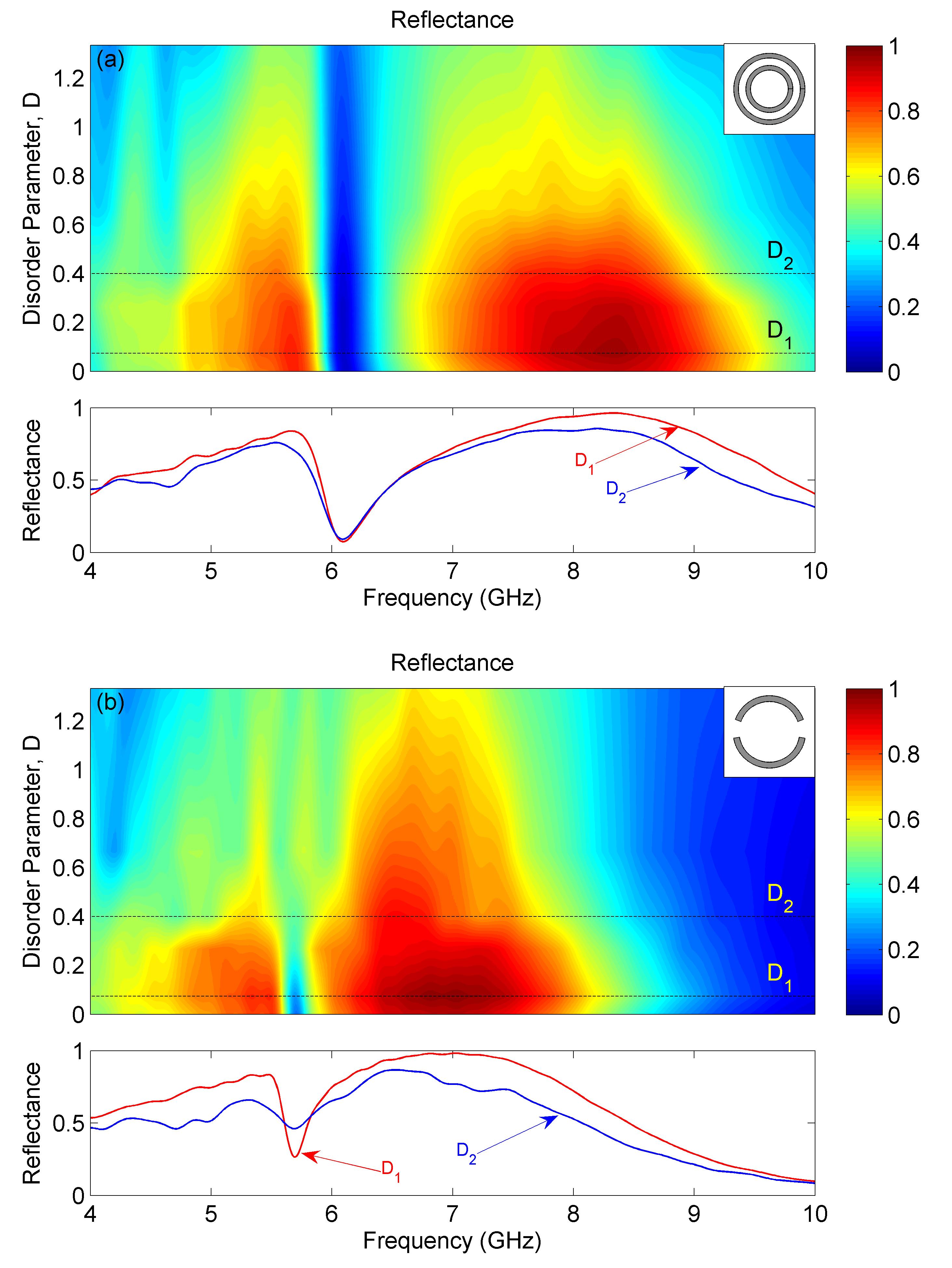}
\caption{Color maps (a) and (b) present the reflectance as a
function of frequency and disorder parameter $D$ for the arrays of
correspondingly concentric and asymmetrically-split rings. The two
marked cross-sections show reflectance profile at $D_1 = 0.07$ and
$D_2 = 0.40$, respectively.}
\end{figure}

The dramatic difference in the behavior of the two metamaterials
becomes apparent, when the dependence of their resonance quality
factors on the disorder is considered (see Fig.~4). Here the
Q-factor represents directly the average strength of the
anti-symmetric component of the current mode induced in the
meta-molecules by the incident electromagnetic wave. If individual
meta-molecules do not interact and thus do not form a collective,
coherent mode, disorder will not affect the spectral lines of the
structure. This is well illustrated by the behavior of the
concentric split ring metamaterial, which we classify as an
"incoherent" metamaterial. On the other hand, if individual
meta-molecules show narrow spectral lines that are affected by
inter-molecular interactions, the spectral lines will
inhomogeneously broaden as a result of randomization of the
inter-molecular distances and thus of the inter-molecular
interactions affecting the lines (spectral shift and split). This is
a trivial case that is not relevant to the "coherent" metamaterial
behavior considered here. In "coherent" metamaterials, narrow lines
are \emph{not seen} in individual meta-molecules and appear only in
ordered arrays. Here randomization leads to radiation damping of the
collective, coherent response of the array. We will see below that
an ensemble of asymmetrically-split rings belongs to this category
of "coherent" metamaterials.
\begin{figure}
\includegraphics[width=80mm]{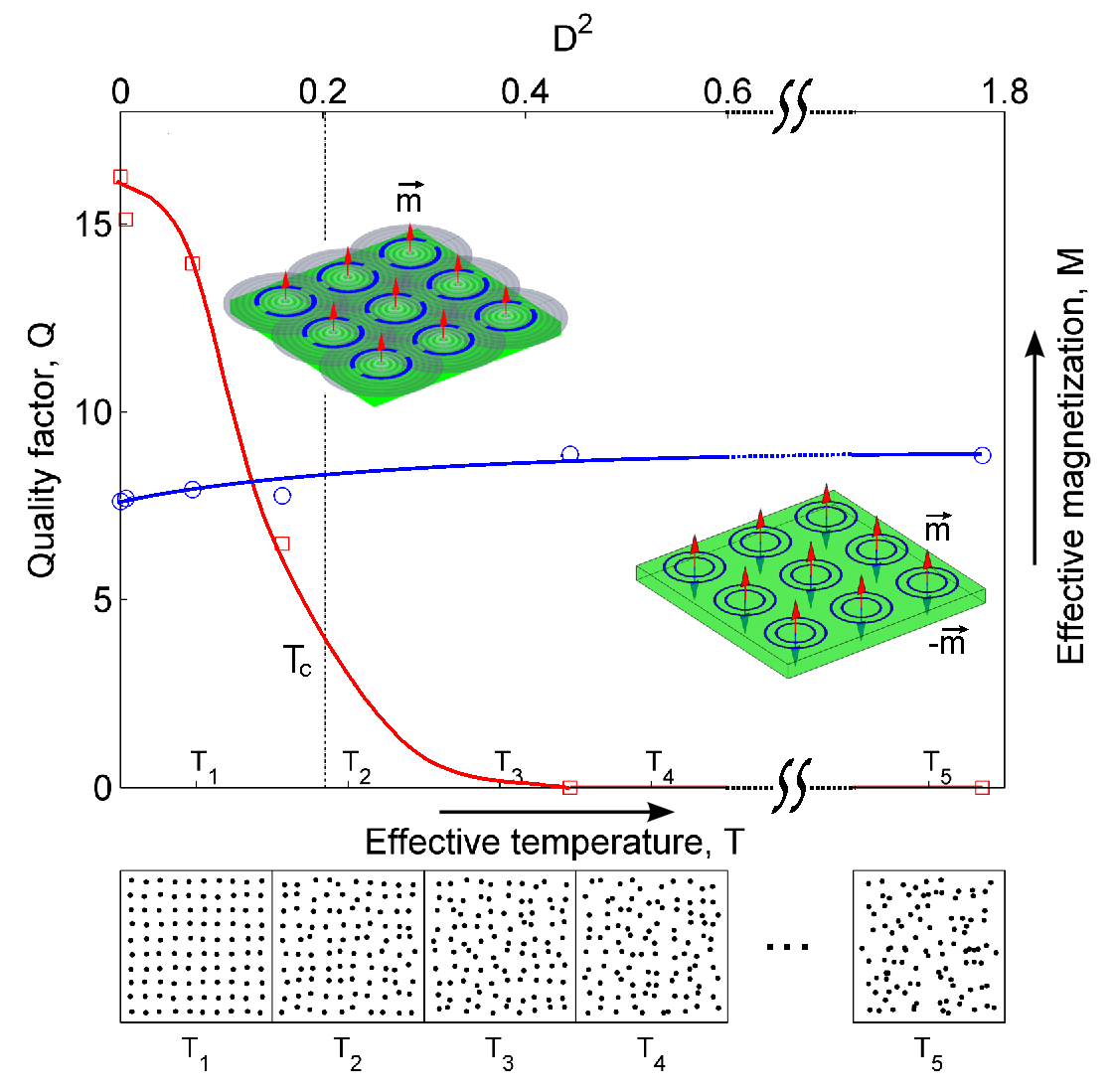}
\caption{Resonance quality factor of the metamaterials' response and
effective magnetization as a function of the disorder parameter
squared (effective temperature) plotted for arrays of concentric
(blue) and asymmetrically-split (red) rings. Points correspond to
experimentally measured values of $Q$, while lines serve as guides
for the reader. The rapid decay of the effective magnetization
associated with the split-rings' response allows the introduction of
a critical temperature $T_c$ similar to the Curie temperature in
ferromagnetics. The insets illustrate the nature of interactions
between the meta-molecules. A sequence of frames with dot arrays
illustrates the increase of disorder in the positions of the
meta-molecules with increasing $D^2$ (effective temperature of the
system).}
\end{figure}

In an ensemble of concentric rings, the Q-factor of the resonant
line is practically independent of disorder. This is due to the fact
that, at resonance, the anti-symmetric currents of the inner and
outer rings create magnetic moments pointing in opposite directions,
thus rendering the total magnetic response of the meta-molecules
extremely weak. Furthermore, by design, the resonant electric
response of the concentring rings is also very weak (which ensures
high Q-factors). Therefore, mutual interactions between the
meta-molecules (both of electric or magnetic dipole-type) are
negligible, and the collective response of the arrays does not
depend on the positions (or disorder) of the molecules, thus
reducing to a direct sum of the individual molecular contributions.
This experimental result is further illustrated by our numerical
simulations presented in Fig.~5, where it is shown that an
anti-symmetric current mode leading to a high Q-factor resonant
response can be excited even in a single isolated concentric-ring
meta-molecule. Negligible magnetization renders the magnetic
behavior of this type of planar metamaterial similar to that of a
\emph{diamagnetic} gas of atoms with paired electrons and thus
represents an example of an ideal \emph{diamagnetic metamaterial.}

On the contrary, in the case of asymmetrically-split ring arrays,
the Q-factor appears to be extremely sensitive to disorder. At the
resonance, the currents induced in the upper and lower arcs of the
split-rings oscillate in opposite phases and therefore create
coherent magnetic dipole moments oscillating in phase. These
magnetic moments are oriented normal to the plane of the array. In a
regular array, interference of waves re-radiated by the oscillating
magnetic dipoles results in the so-called magneto-inductive waves
\cite{magnind}, which are confined to the plane of the array
 mediating efficiently strong interactions between the meta-molecules
(see illustration in the inset to Fig.~4). The increase of disorder
leads to intense scattering of these waves into free space
(conversion into free space electromagnetic waves) and,
consequently, to losses that reduce the strength of the
anti-symmetric current mode. Consequently, the inter-molecular
interactions become weaker resulting in a lower Q-factor and
vanishing net magnetization. Importantly, in the limiting case of a
single isolated asymmetrically-split ring, the magnetic dipole
scattering is at maximum and the high-Q resonance is completely
absent, as opposed to a pair of concentric rings (see Fig.~5). Thus,
the broadening/weakening of the collective response cannot be
explained by splitting or shifting of the high-Q resonances of the
individual meta-molecules that form clusters of closely spaced or
even overlapping split rings in the disordered arrays, as the
resonance simply does not exist for an individual meta-molecule.
Note, that the Q-factor of the metamaterial's response here is not
only a direct measure of the strength of the anti-symmetric current
mode, but also a measure of the metamaterial's effective
magnetization, induced normal to the plane of the array by the
incident electromagnetic wave.

Intriguingly, the rapid decay of the effective magnetization with
increasing disorder resembles phase transitions in magnetic
solid-state systems, such as the behavior of a two-dimensional array
of spins (Ising ferromagnet \cite{ising}) in the absence of external
fields with increasing temperature. This similarity is not
coincidental and reflects a common physical origin. In the Ising
model the Hamiltonian of the system is given by $H=\Sigma J_{ij}
S_i~S_j$, where $J_{ij}$ is a constant that reflects the strength of
interaction between spins $S_i$ and $S_j$. With increasing
temperature, the ratio of the interaction energy $J$ over the energy
of thermal fluctuations ($k_\text{B} T$) decreases. This leads to a
second-order phase transition from the initial ferromagnetic ordered
state, where all the spins pointed in the same direction, to a
diamagnetic disordered state with random spin orientations,
corresponding to a vanishing macroscopic magnetization. In the
ensemble of asymmetrically-split rings, the \emph{induced} magnetic
moments $m$ of the individual meta-molecules interact in a similar
way and therefore the system's Hamiltonian will have the following
form: $H=<\Sigma j_{ij} m_i~m_j>$, where brackets indicate time
average over one period of the incident wave and no coupling exists
between the oscillating magnetic dipoles and external magnetic field
due to their orthogonality. Here the transition to the diamagnetic
state results from the destruction of the coherent surface state by
scattering. Furthermore, by considering random displacements of
meta-molecules in the metamaterial arrays as a result of 'heating'
(similar to real crystalline solids) the disorder parameter can be
related to an effective temperature as follows, $T \sim D^2$.
Although the analogy between the asymmetrically-split ring system
and Ising ferromagnet is phenomenological, direct mapping into more
complicated spin systems, such as spin glasses \cite{spinglass} is
also possible.
\begin{figure}[hb]
\includegraphics[width=80mm]{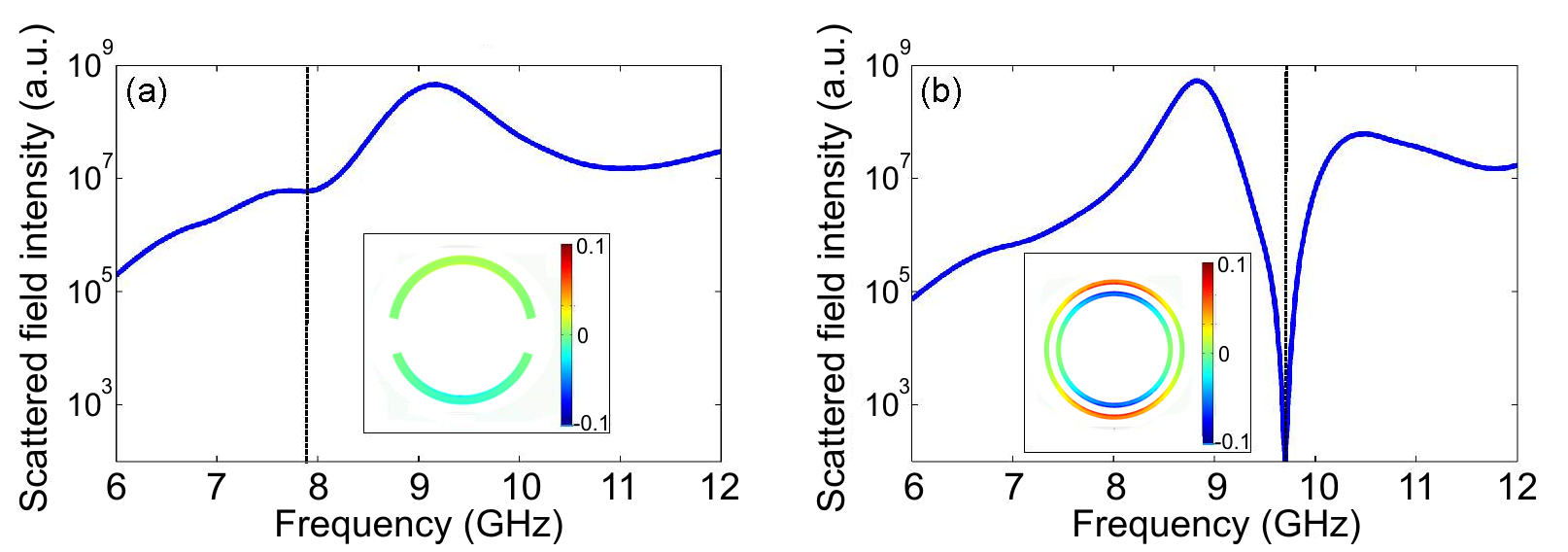}
\label{figS1} \caption{(a) Scattered intensity for a single
asymmetrically split ring (ASR) and (b) for a single concentric ring
structure. The dashed lines mark the position of the antisymmetric
current configuration for each case. Color maps of the associated
current modes (projection along the horizontal axis) are provided in
the insets.}
\end{figure}

The loss of coherency by disorder presented here is fundamentally
different from the recently investigated inhomogeneous line
broadening in disordered photonics crystals\cite{pcs} and
metamaterials \cite{kivshar06,dsmith} that were concerned with
negative permeability associated with "incoherent" response due to
random changes in the geometry of the individual meta-molecules. On
the contrary, in the "coherent" metamaterial described here, the
sharp spectral feature is a collective resonance of the entire
ordered structure. This behavior is reminiscent of another well
known phenomenon in many-body physics, namely the M\"{o}ssbauer
effect \cite{mossb}, where transitions of atomic nuclei with
extremely narrow line-widths can only be observed when the atoms
form a crystal lattice, while in an atomic gas the emission line is
strongly broadened by the recoil during the emission of the high
energy gamma quanta. In a similar way, a single meta-molecule of the
coherent metamaterial does not exhibit the narrow resonance
associated with the antisymmetric current mode (also known as
trapped-mode resonance) as a result of scattering losses, while in a
regular array no scattering losses are possible for wavelengths
longer than the array pitch and thus a low-frequency high quality
mode is formed.

The two planar metamaterial structures considered above are examples
of artificial media with strong and weak inter-meta-molecular
interactions representing two characteristic and antipode classes of
what we call "coherent" and "incoherent" metamaterials. The
fundamental difference in the nature of their narrow resonances,
collective in the first case and individual in the second,
determines the potential applications of these structures, most
notably the lasing spaser \cite{spaser}. The lasing spaser, a
metamaterial analog of spaser \cite{stockman} is a planar
narrow-diversion coherent source of electromagnetic radiation that
is fuelled by plasmonic oscillations of a two-dimensional resonator
array. Here the coherency of the optical source is ensured by the
synchronous oscillations of the plasmonic currents in the array. In
a "coherent" metamaterial formed by asymmetrically-split rings, the
regular array gives the highest value of quality factor compared to
disordered arrays. We argue that, similarly, when both phased
(coherent) and uncorrelated (incoherent) current oscillations are
present in an array of meta-molecules of this type, the uncorrelated
component will decay more rapidly. Therefore in the presence of gain
the phased, coherent component of amplified spontaneous current
fluctuations will win over incoherent fluctuations providing for a
self-starting regime of the lasing spaser. On the contrary,
"incoherent" metamaterials with weak inter-molecular interactions,
such as concentric rings, do not possess a mechanism of
synchronization of current oscillation in individual meta-molecules
and are not suitable for lasing spaser applications. However,the
resonant properties of "incoherent" metamaterials are more tolerant
to disorder making them more suitable for manufacturing using
methods prawn to imperfections such as self-assembly
\cite{dorota}.

Finally, another intriguing and unique property of the "coherent"
metamaterials will be the dependence of their transmission and
reflection spectra on the size of the sample array. As has been
demonstrated by numerical calculations narrow band resonance is only
featured in the spectrum when a large number of meta-molecules are
involved in the formation of the optical response. We argue that in
a similar way optical spectra measured with spatially coherent and
incoherent sources of electromagnetic radiation will be different,
with sharp resonances featured only under coherent illumination.

 The authors are grateful to Peter Nordlander for stimulating discussions
and guidance on the electromagnetic response of disordered systems.
This work is supported by the European Union through the FP7
ENSEMBLE project and by the Engineering and Physical Sciences
Research Council (UK) through the Nanophotonics Portfolio Grant and
the International Collaborative grant with National Taiwan
University.

\end{document}